# ON THE INFORMATION ENGINE OF CIRCUIT DESIGN

*Denis V. Popel*

Department of Computer Science,
Baker University, Baldwin City, KS 66006,
popel@ieee.org

*Nawar Al Hakeem*

Department of Computer Science,
University of Wollongong (Australia), Dubai Campus, U.A.E.,
hakeem@ieee.org

**ABSTRACT**

This paper addresses a new approach to find a spectrum of information measures for the process of digital circuit synthesis. We consider the problem from the *information engine* point of view. The circuit synthesis as a whole and different steps of the design process (an example of decision diagram is given) are presented via such measurements as *entropy*, *logical work* and *information vitality*. We also introduce new information measures to provide better estimates of synthesis criteria. We show that the basic properties of information engine, such as the conservation law of information flow and the equilibrium law of information can be formulated.

## 1. INTRODUCTION

For three decades history of application of information theory methods to the design of integrated circuits, many promising results have been obtained in the following fields:

- *machine learning* [7];
- *entropy based minimization* [5];
- *testing* of digital circuits [1];
- *estimation of power dissipation* [4];
- other related design problems (influence of types of logical gates on entropy [2], estimation of logical work in networks and programs [3]).

All previous results constitute that in information networks, of which digital circuits are an integral component, one deals with a variety of logical operations and complex processes delivering signals before information reaches the final destination. These information systems seem to need a new basic theory which helps to understand the system in its entirety in order to provide a basis for designing the total system. Such a theory was introduced in [10], and further developed in [11]. In accordance with the theory, the machines which deal with information can be generally referred to as *information engines* which model the actual information systems. In this paper, we propose such an interpretation for circuit design which considers behaviour of circuit synthesis in the context of information measures [3, 9]. Hence, entropy is used as a complexity characteristic in the circuit-design optimization process and generally in signal processing [6]. Altogether, we describe a new method of analysis for geometry-based synthesis in which the main contribution is using information theoretic measures for calculating parameters of geometry in gate-level circuit design.

## 2. PRELIMINARIES AND ASSUMPTIONS

Consider logic representation of a digital circuit in the form of *Boolean function* $f$ treated as the mapping $\{0,1\}^n \to \{0,1\}^m$ over the variable set $X = \{x_1, \cdots, x_n\}$. Here, $n$ is the number of variables (inputs), and $m$ is the number of functions (outputs).

### 2.1. Information Theoretic Notations

In order to quantify the content of information for a finite field of events $A = \{a_1, a_2, \cdots, a_n\}$ with probabilities distribution $\{p(a_i)\}$, $i = 1, 2, \cdots, n$, Shannon introduced the concept of *entropy* [9]: $H(A) = -\sum_{i=1}^{n} p(a_i) \cdot \log p(a_i)$, where $\log$ denotes the base 2 logarithm. For two finite fields of events $A$ and $B$ with probability distribution $\{p(a_i)\}, i = 1, 2, \cdots, n$, and $\{p(b_j)\}$, $j = 1, 2, \cdots, m$, probability of the joint occurrence of $a_i$ and $b_j$ is joint probability $p(a_i, b_j)$, and there is conditional probability, $p(a_i|b_j) = p(a_i, b_j)/p(b_j)$. The *conditional entropy* of $A$ given $B$ is defined by $H(A|B) = -\sum_{i=1}^{n} \sum_{j=1}^{m} p(a_i, b_j) \cdot \log p(a_i|b_j)$.

In case of logic networks and signal processing, we assume that the sets of values of a function $f$ and arbitrary variable $x$ are two finite fields [5].

### 2.2. Gate-level Information Measuring

Let us consider simple gates and their information content in order to analyze the information transmission through each gate. The concept of *logical work* is crucial for information measures of Boolean networks [3]. We use the following notation for each gate: the input entropy $H(X)$, and the output entropy of function $f$ is $H(f)$.

Table 1: Information equivalent of primitive gates in bits

| Library gates | $\mathcal{NOT}$ | $\mathcal{AND}$ | $\mathcal{OR}$ | $\mathcal{EXOR}$ |
|---|---|---|---|---|
| | | | | |
| Function $f$ | [01] | [0001] | [0111] | [0110] |
| Maximum input information $H(X)$ | 1.0 | 2.0 | 2.0 | 2.0 |
| Output information $H(f)$ | 1.0 | 0.81 | 0.81 | 1.0 |
| Gate information measure $H(X){-}H(f)$ | 0.0 | 1.19 | 1.19 | 1.0 |
| Transmission of information $H(f\|x)$ | 1.0 | 0.5 | 0.5 | 1.0 |

**Definition 1** *The conditional entropy $H(f|x)$ between arbitrary variable $x$ and the function $f$ is a measure of transmission of information from input $x$ to the output $f$.*

**Definition 2** *The difference between input and output entropies of a gate,*

$$I_{gate} = H(X) - H(f),$$

*is an information theoretic measure of the $gate$.*

**Example 1** *For the function $\mathcal{AND}$ we obtain the following entropy measures:* $H(X) = -{}^1/_4 \cdot \log {}^1/_4 - {}^1/_4 \cdot \log {}^1/_4 - {}^1/_4 \cdot \log {}^1/_4 - {}^1/_4 \cdot \log {}^1/_4 = 2$ bits, $H(f) = {}^1/_4 \cdot \log {}^1/_4 - {}^3/_4 \cdot \log {}^3/_4 = 0.81$ bits, $I_{gate} = 2 - 0.81 = 1.19$ bits, $H_{trans}(f|x_1) = H_{trans}(f|x_2) = 0.5$ bits.

The information measures of primitive gates are given in Table 1. Note, the uncertainty of the output of a digital network is never increased by knowledge of input patterns. In other words, the digital network does not increase information carried by the inputs.

## 3. INFORMATION ENGINE OF CIRCUIT SYNTHESIS

Consider an information engine with one input port and one output port, where the initial description of the logic function $f$ is converted to output description. Any process done by the information engine can be considered as a composition of two types of processes, *isentropic* process and entropy changing process. Successful (100% functionality) design is an isentropic process that does not change the function itself. When we deal with an entropy changing process that requires the conversion of input and output combinations or ensembles, we call such a process *iso-vitality*.

The concept of information engine of gate-level circuit design is based on the following notation. For every input description of the logic function $f$, there exists the logical *NetWork* (*NW*) which converts input ensemble to output ensemble. Hence, the loss of uncertainty or network information can be characterized by $I_{NW} = H_{loss}(f|X) = H(X) - H(f)$. For each input $x_i$ of the logical network, the value of information transmitted to the output $f$ is $H_{trans}(x_i) = H(f|x_i)$.

### 3.1. Circuit Design as Isentropic Process

Consider a circuit design process as a sequence of steps which operate with input variable $X$, and correspond to network representation of the function $f$. Let $q(f;NW)$ be the *logical work* required to convert initial description of $f$ to the network $NW$ via an information engine. The concept of logical work for digital networks was also discussed in [3]. The circuit design process is reversible, i.e., it is possible to produce the initial function description from the network, which means it is an isentropic process: $H(f) = I(f;NW)$ and $I(X;NW) = H(X)$.

For an isentropic process of circuit design, we can use the information potential measure defined next.

**Definition 3** *Information Potential is the inferior bound of quantity of logical work $q(f;NW)$ under all possible $NW$ design processes,*

$$Q(f;NW) = Inf[q(f;NW)] \ for\ all\ possible\ NW$$

*This is the information potential of logic function $f$ with respect to network $NW$.*

Another important information concept is,

**Definition 4** *Information Vitality is the information potential of an ensemble $X$ with respect to network $NW$ per unit of information,*

$$T(NW) = Q(f;NW)/H(f)$$

*This is the information vitality equation.*

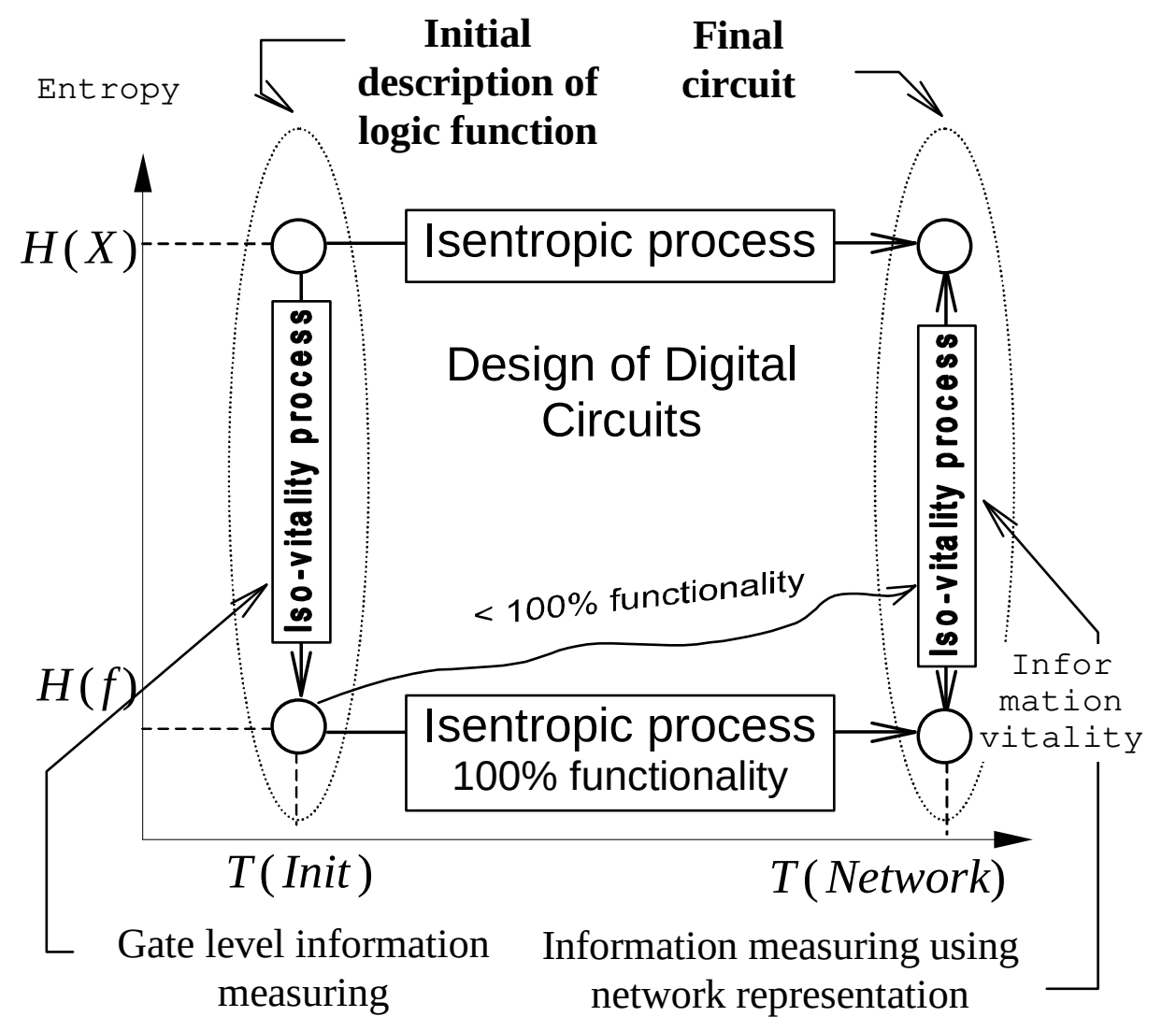

Figure 1: H-T-Diagram of gate-level circuit design process

The information potential characterizes the cost of final realization of the logic function $f$, therefore, the search of the inferior bound of $q(f;NW)$ represents a search of the logic function that will minimize the cost of realization.

The dynamic behaviour of an information engine of gate-level circuit design can be best described in the entropy vs. vitality diagram, so called H-T-Diagram (Figure 1). The design of digital circuit is an isentropic process with respect to inputs and outputs. Information verification of circuit design is an iso-vitality process. We can evaluate the loss of information via construction of a logic network for the verified function. The synthesis of an information engine for gate-level circuit design deals with all parts of H-T-Diagram in addition to information measuring which is an essential part of the synthesis process.

### 3.2. Circuit Layout Measuring

We focus in our study on a design process for a specified geometry style defined over a *library* $\mathcal{L}$ of gates $\mathcal{G}_1, \ldots, \mathcal{G}_t$, where $t$ is a total number of gates in the library. We consider geometry $p \times q$ as one of the main components of the design specification. Note that geometry incorporates many parameters: the maximal number of $p$ levels, the maximal number $q$ of gates in each level of the network, types of logic gates from the standard logic library $\mathcal{L}$, network architecture (permissible interconnections between cells, levels, inputs and outputs of the circuit). It is essential to consider different geometry realizations, and various schemes of assigning logic elements to different circuit levels (see, for example, $\mathcal{AND}-\mathcal{OR}-\mathcal{EXOR}$ networks in [8]).

The main idea of information measuring of a geometry is to give numerical estimations and establish propositions for the design parameters. Here we define *information capacity* for a gate, a library of gates, and geometry.

We select gates from the cell library $\mathcal{L}$. The information equivalents of library gates are given in Table 1. We consider $I_{gate}$ as an *information capacity* of the gate. Here, we do not take into account the information carried by inputs and outputs of the network, because such an information does not influence geometry parameters.

**Definition 5** Information capacity $I_{\mathcal{L}}$ of a library $\mathcal{L}$ *of gates* $\mathcal{G}_1, \ldots, \mathcal{G}_t$ *is the entire amount of information capacities of the gates:*

$$I_{\mathcal{L}} = \textstyle\sum_{\mathcal{L}} I_{gate}.$$

**Example 2** *Given the gate library* $\mathcal{L} = \{\mathcal{NOT}, \mathcal{AND}, \mathcal{OR}\}$, *information capacity is* $I_{\mathcal{L}} = I_{gate}$ *(*$\mathcal{NOT}$*)* $+I_{gate}$ *(*$\mathcal{AND}$*)* $+I_{gate}$ *(*$\mathcal{OR}$*)* $= 2.38\ bit$.

**Definition 6** Information capacity $I_{\mathcal{G}}$ of a single cell *is the maximal information capacity of all gates from the given (fixed) cell library* $\mathcal{L}$*:*

$$I_{\mathcal{G}} = max\{I_{gate}\}\ for\ all\ gates\ from\ library\ \mathcal{L}.$$

Based on these definitions, we can assume that information capacity of $p \times q$ geometry over a fixed library $\mathcal{L}$ of gates $\mathcal{G}_1,\ldots,\mathcal{G}_t$ is a composition of information capacities of the cells:

$$I_{Geometry} = p \cdot q \cdot I_{\mathcal{G}}$$

**Example 3** *Given the cell library* $\mathcal{L} = \{\mathcal{NOT}, \mathcal{AND}, \mathcal{OR}\}$ *and a geometry of* $2 \times 2$, *the information capacity of cells in the first level is* $I_{\mathcal{G}} = 1.19\ bit$, *and in the second level is* $I_{\mathcal{G}} = 0.595\ bit$. *The information capacity of the geometry is equal to* $I_{Geometry} = 3.57\ bit$.

We summarize the information measures for different specifications in Table 2. These measures allow us to make a priori decision about the efficiency of a geometry.

## 4. INFORMATION ENGINE OF DECISION DIAGRAM DESIGN

Here we apply the given above approach to the design process of Decision Diagrams (DDs). DDs are widely used graph structures for representation and manipulation of logic functions in circuit synthesis [8]. The concept of information engine of DD design is based on the following notations. For every input description of the logic function $f$, there exist the logical network which converts input ensemble to output ensemble.

Table 2: Information capacity of geometries

| | Geometry | | |
|---|---|---|---|
| | $\{\mathcal{NOT},$ $\mathcal{AND},\mathcal{OR}\}$ | $\{\mathcal{NOT},$ $\mathcal{EXOR}\}$ | $\{\mathcal{NOT},\mathcal{AND},$ $\mathcal{OR},\mathcal{EXOR}\}$ |
| | $2 \times 2$ | | |
| $I_{Geometry}$ | 3.57 | 3.00 | 3.57 |
| | $3 \times 3$ | | |
| $I_{Geometry}$ | 6.2475 | 5.2500 | 6.2475 |

Consider the design process as a sequence of steps which operates with variables from $X$ and correspond to an expansion of the function $f$ [8]. The DD design is reversible, i.e. it is possible to produce the initial function description from DD, in other words, it is isentropic process: $H(f) = I(f; DD)$ and $I(X; DD) = H(X)$. The dynamic behaviour of an information engine of DD design can be outlined in entropy vs. vitality diagram (Figure 1).

We use two information measures: conditional entropy $H(f|DD)$ and mutual information $I(f; DD)$, to describe DD design process. The *initial state* of this process (no DD exists) is characterized by the maximal value of the conditional entropy $H(f|DD) = H(f)$, where $DD = \{\oslash, \oslash\}$. Nodes are recursively attached to DD according top-to-down strategy. The entropy $H(f|DD)$ is reduced, information $I(f; DD)$ is increased because of each attached couple $(x, \omega)$ contributes its portion of information about the function. Any *intermediate state* can be described in terms of information theory by equation

$$I(f; DD) = H(f) - H(f|DD). \quad (1)$$

We maximize the information $I(f; DD)$ that corresponds to minimization of entropy $H(f|DD)$, on each step of decision diagram design.

The *final state* of DD design is characterized by $H(f|DD) = 0$ and $I(f; DD) = H(f)$, i.e. $DD$ represents logic function $f$.

## 5. CONCLUDING REMARKS

It has been shown that information theory measures of the circuit design process are useful and give new possibilities to improve the efficiency of the recently developed techniques. The main contribution of this paper is the inclusion of circuit information content in the design process and signal propagation in the context of an information engine. The extension of the recently developed technique of circuit design includes information quantification of a cell library, estimation of circuit geometry, and measuring decision diagram design. Hence, results obtained with the proposed technique show that it can be useful for a priori analysis of circuit synthesis.